\documentclass[11pt,a4paper]{article}
\usepackage{amsfonts,amsmath,amssymb,accents,amsthm}
\usepackage{verbatim}

\usepackage{hyperref}
\usepackage[normalem]{ulem}

\usepackage{tikz} 
\usetikzlibrary{calc,arrows,positioning,fit,petri}

\newcommand{\dis}{\displaystyle}

\newcommand{\noi}{\noindent}
\newcommand{\halmos}{\rule{1ex}{1.4ex}}
\newcommand{\QED}{\nopagebreak{\hspace*{\fill}$\halmos$\medskip}}

\newcommand{\quand}{\quad\mbox{and}\quad}

\newtheoremstyle{mythm}
  {}
  {}
  {\itshape}
  {}
  {\bfseries}
  {}
  {.5em}
  {#1 #2 \thmnote{(#3)}}

\theoremstyle{mythm}
\newtheorem{theorem}{Theorem}
\newtheorem{proposition}[theorem]{Proposition}
\newtheorem{lemma}[theorem]{Lemma}
\newtheorem{exercise}[theorem]{Exercise}
\newtheorem{corollary}[theorem]{Corollary}
\newtheorem{conjecture}[theorem]{Conjecture}

\newtheorem{counterex}[theorem]{Counterexample}

\newcommand{\bt}{\begin{theorem}}
\newcommand{\et}{\end{theorem}}
\newcommand{\bl}{\begin{lemma}}
\newcommand{\el}{\end{lemma}}
\newcommand{\bp}{\begin{proposition}}
\newcommand{\ep}{\end{proposition}}
\newcommand{\bcor}{\begin{corollary}}
\newcommand{\ecor}{\end{corollary}}
\newcommand{\br}{\begin{remark}\rm}
\newcommand{\er}{\end{remark}}
\newcommand{\bcon}{\begin{conjecture}}
\newcommand{\econ}{\end{conjecture}}
\newcommand{\bex}{\begin{exercise}}
\newcommand{\eex}{\end{exercise}}
\newcommand{\bcou}{\begin{counterex}}
\newcommand{\ecou}{\end{counterex}}

\newenvironment{Proof}[1][]{\noi\textbf{Proof #1}}{\QED}
\newcommand{\bpro}{\begin{Proof}}
\newcommand{\epro}{\end{Proof}}

\newcommand{\be}{\begin{equation}}
\newcommand{\ee}{\end{equation}}
\newcommand{\ba}{\begin{array}}
\newcommand{\ea}{\end{array}}
\newcommand{\bac}{\begin{array}{r@{\,}c@{\,}l}}
\newcommand{\bc}{\be\begin{array}{r@{\,}c@{\,}l}}
\newcommand{\ec}{\end{array}\ee}

\newcommand{\al}{\alpha}

\newcommand{\de}{\delta}

\newcommand{\eps}{\varepsilon}
\newcommand{\la}{\lambda}

\newcommand{\sig}{\sigma}

\newcommand{\Si}{{\cal S}}

\newcommand{\Xc}{{\cal X}}

\newcommand{\R}{{\mathbb R}}

\newcommand{\Z}{{\mathbb Z}}

\renewcommand{\P}{{\mathbb P}}

\newcommand{\up}{\uparrow}
\newcommand{\down}{\downarrow}
\newcommand{\sub}{\subset}

\newcommand{\ti}{\tilde}

\newcommand{\ov}{\overline}

\newcommand{\di}{\mathrm{d}}

\begin{document}

\makeatletter\@addtoreset{equation}{section}
\makeatother\def\theequation{\thesection.\arabic{equation}} 

\renewcommand{\labelenumi}{{\rm (\roman{enumi})}}
\renewcommand{\theenumi}{\roman{enumi}}

\title{How much market making does a market need?}
\author{V\'it Per\v{z}ina\footnote{Univerzita Karlova,
 Matematicko-fyzik\'aln\'i fakulta, Ke Karlovu 3, 121 16 Praha 2,
 Czech Republic; perzina@gmail.com}
\and Jan~M.~Swart\footnote{The Czech Academy of Sciences,
 Institute of Information Theory and Automation,
 Pod vod\'arenskou v\v e\v z\' i 4,
 182 08 Praha 8,
 Czech Republic;
 swart@utia.cas.cz}
}

\date{\today}

\maketitle

\begin{abstract}\noi
We consider a simple model for the evolution of a limit order book in which
limit orders of unit size arrive according to independent Poisson processes.
The frequencies of buy limit orders below a given price level, respectively sell
limit orders above a given level are described by fixed demand and supply
functions. Buy (resp.\ sell) limit orders that arrive above (resp.\ below) the
current ask (resp.\ bid) price are converted into market orders. There is no
cancellation of limit orders. This model has independently been reinvented by
several authors, including Stigler in 1964 and Luckock in 2003, who was able
to calculate the equilibrium distribution of the bid and ask prices. We extend
the model by introducing market makers that simultaneously place both a buy
and sell limit order at the current bid and ask price. We show how the
introduction of market makers reduces the spread, which in the original model
is unrealistically large. In particular, we are able to calculate the exact
rate at which market makers need to place orders in order to close the spread
completely. If this rate is exceeded, we show that the price settles at a
random level that in general does not correspond the Walrasian equilibrium
price.
\end{abstract}
\vspace{.5cm}

\noi
%
{\it Acknowledgments.} Work sponsored by GA\v{C}R grant 15-08819S.

{\setlength{\parskip}{-2pt}\tableofcontents}

\newpage

\section{Introduction}

\subsection{Informal description of the model}

We will be interested in a simple mathematical model for the evolution of a
limit order book, as used on a stock market or commodity market. The basic
model we are interested in has been independently (re-)invented at least four
times, by \cite{Sti64,Luc03,Pla11,Yud12b}. The aim of the model is not
so much to identify optimal strategies for traders, but rather to identify,
in a simplified set-up, the basic mechanisms that lie behind the observed
shape and time evolution of real order books.

Even in regard to this modest aim, the original model as first formulated in
\cite{Luc03} is not particularly succesful. Indeed, it leads to a highly
unrealistic order book, in which the spread is very large, while far from the
equilibrium price the number of limit orders grows without bounds. In the
present paper we propose an extension of the model that fixes one unrealistic
aspect of the original model, by closing the spread (at least for a special
choice of the parameters), but retains other unrealistic
features. Nevertheless, it is hoped that by identifying the basic mechanisms
that lie behind the behavior of simple models, eventually a more realistic
model can be developed that leads to a better understanding of the mechanisms
that shape real order books.

Since our aim is not to identify trading strategies, we allow traders to
behave in a way that can be far from their optimal strategy, which in a
setting where time is continuous and trading is open ended may anyway be hard
to identify. Also, we do not identify individual traders, i.e., we allow for
the possibility that some of the orders arriving at different times may in
fact be placed by one and the same trader, but do not record this information.

Our starting point is the model as first formulated in full generality in
\cite{Luc03}. In this model, limit orders of unit size arrive according to
independent Poisson processes. The frequencies of buy limit orders below a
given price level, respectively sell limit orders above a given level are
described by fixed demand and supply functions. Buy (resp.\ sell) limit orders
that arrive above (resp.\ below) the current ask (resp.\ bid) price are
converted into market orders. There is no cancellation of limit orders.
Following \cite{Swa16}, we add a second type of traders, who always place
market orders, regardless of the current price levels. From a modeling point
of view, we can view these orders as buy (resp.\ sell) that arrive at such a
high (resp.\ low) prices that they are always converted into market orders,
except when there are currently no matching sell (resp.\ buy) limit orders in
the order book. From a mathematical point of view, the addition of this kind
of orders is useful since it allows for positive recurrent behavior, which
is never possible in the original model.

The novelty of the present paper kies in the introduction a new type of
trader, who is a market maker or more general any liquidity supplier, who
instead of only buying or selling does both, with the aim of making a profit
from the spread. We model the effect of such market makers by saying that
according to a fixed Poisson rate, a buy and sell limit order of unit size are
simultaneously placed at the current bid and ask prices.

In Section~\ref{S:station}, we adapt the method of Luckock \cite{Luc03} for
calculating the spread to the generalized model (Theorem~\ref{T:Luckock2}) and
show that the introduction of market makers reduces the spread, until it
closes completely if the rate at which market makers place orders equals the
Walrasian volume of trade. In Section~\ref{S:freeze} we show that if the rate
of market making is increased beyond this point, then the bid and ask prices
converge to a random limit that does not need to correspond to the Walrasian
equilibrium price (Theorem~\ref{T:freeze}).

In the remainder of this introduction, we formulate our model precisely and
settle notation (Subsection~\ref{S:model}) and discuss its history
(Subsection~\ref{S:hist}). Section~\ref{S:orig} is devoted to the original
model due to Stigler and Luckock while Section~\ref{S:market} discusses the
new phenomena due to the introduction of market making.

\subsection{Definition of the model}\label{S:model}

Let $I=(I_-,I_+)\sub\R$ be a nonempty open interval, modelling the possible
prices of limit orders, and let $\ov I=[I_-,I_+]\sub[-\infty,\infty]$ denote
its closure. Recall that a \emph{counting measure} on $I$ is
a measure $\mu$ that can be written as a countable sum of delta measures. At any
given time, we represent the state of the order book by a pair $(\Xc^-,\Xc^+)$
of counting measures on $I$, where we interpret the delta measures that $\Xc^-$
(resp.\ $\Xc^+$) is composed of as buy (resp.\ sell) limit orders of unit size
at a given price. We assume that:
\begin{enumerate}
\item there are no $x,y\in I$ such that $x\leq y$ and $\Xc^+(\{x\})>0$,
  $\Xc^-(\{y\})>0$,
\item $\Xc^-\big([x,I_+)\big)<\infty$ and $\Xc^+\big(I_-,x]\big)<\infty$ for
  all $x\in I$.
\end{enumerate}
Here, the first condition says that the order book cannot simultaneously
contain a buy and sell limit order when the ask price of the seller is lower
than or equal to the bid price of the buyer. The second condition guarantees
that
\bc
\dis M^-&:=&\dis\max\big(\{I_-\}\cup\{x\in I:\Xc^-(\{x\})>0\}\big),\\[5pt]
\dis M^+&:=&\dis\min\big(\{I_+\}\cup\{x\in I:\Xc^+(\{x\})>0\}\big),
\ec
are well-defined, which can be interpreted as the current bid and ask prices.
Note that $M^\pm:=I_\pm$ if the order book contains no limit orders of the
given type. It is often convenient to represent the order book by the signed
counting measure $\Xc:=\Xc^+-\Xc^-$. We let $\Si_{\rm ord}$ denote the space
of all signed measures of this form, with $\Xc^-$ and $\Xc^+$ satisfying the
conditions (i) and (ii) above.

The dynamics of the model are described by two functions $\la_\pm:\ov I\to\R$,
which we call the \emph{demand function} $\la_-$ and \emph{supply function}
$\la_+$, and a nonnegative constant $\rho\geq 0$, which will represent the
\emph{rate of market makers}. We assume that:
\begin{itemize}
\item[{\rm(A1)}] $\la_-$ is nonincreasing, $\la_+$ is nondecreasing,
\item[{\rm(A2)}] $\la_\pm$ are continuous functions,
\item[{\rm(A3)}] $\la_+-\la_-$ is strictly increasing,
\item[{\rm(A4)}] $\la_\pm>0$ on $I$.
\end{itemize}
We let $\di\la_\pm$ denote the measures on $I$ defined by
$\di\la_\pm\big([x,y]\big):=\la_\pm(y)-\la_\pm(x)$ $(x,y\in I,\ x\leq y)$.
In particular, $\di\la_-$ is a negative measure and $\di\la_+$ is a positive
measure. We consider a continuous-time Markov process $(\Xc_t)_{t\geq 0}$
that takes values in the space $\Si_{\rm ord}$ and whose dynamics have the
following description.
\begin{itemize}
\item[] \textbf{Buy orders inside the interval} With Poisson local rate
  $-\di\la_-$, a trader comes and places a buy limit order at a price $x$, or
  takes the best available sell limit order at a price $\leq x$, if there is
  one, i.e., $\Xc\mapsto\Xc-\de_{x\wedge M^+}$.
\item[] \textbf{Buy orders outside the interval} With Poisson rate
  $\la_-(I_+)$, a trader comes and takes the best available sell limit order,
  if there is one, i.e., $\Xc\mapsto\Xc-\de_{M^+}$ if $M^+<I_+$ and nothing
  happens otherwise.
\item[] \textbf{Sell orders inside the interval} With Poisson local rate
  $\di\la_+$, a trader comes and places a sell limit order at a price $x$, or
  takes the best available buy limit order at a price $\geq x$, if there is
  one, i.e., $\Xc\mapsto\Xc+\de_{x\vee M^-}$.
\item[] \textbf{Sell orders outside the interval} With Poisson rate
  $\la_+(I_-)$, a trader comes and takes the best available buy limit order,
  if there is one, i.e., $\Xc\mapsto\Xc+\de_{M^-}$ if $M^->I_-$ and nothing
  happens otherwise.
\item[] \textbf{Market makers} With Poisson rate $\rho$, a market maker
  arrives who places both a buy and sell limit order at the current ask and
  bid prices, provided these lie inside $I$, i.e.,
  $\Xc\mapsto\Xc-1_{\{M^->I_-\}}\de_{M^-}+1_{\{M^+<I_+\}}\de_{M^+}$.
\end{itemize}
Here, the phrase ``with Poisson local rate $\di\la_+$'' means that sell limit
orders with prices inside some measurable set $A\sub I$ arrive with Poisson
rate $\di\la_+(A)$, which is independent for disjoint sets $A$. We assume
that all Poisson processes governing different mechanisms (buy/sell market/limit
orders, and market makers) are independent. After \cite{Sti64,Luc03}, we call
the Markov process $(\Xc_t)_{t\geq 0}$ the \emph{Stigler-Luckock model} with
demand and supply functions $\la_\pm$ and rate of market makers $\rho$.

We make the assumptions (A2)--(A4) for technical simplicity. As explained in
Appendix A.1 of \cite{Swa16}, these assumptions can basically be made without
loss of generality. In particular, models for which (A2) and (A3) fail can be
obtained as functions of models for which (A2) and (A3) hold. In particular,
this applies to discrete models in which limit orders can only be placed at
integer prices. To explain this on a concrete example, consider a model with a
price interval of the form $I=(0,2n)$ where $n\geq 1$ is some integer, and
demand and supply functions that satisfy
\be\label{evenodd}
\di\la_-=-1_{\{\lceil x\rceil\mbox{ is even}\}}\di x
\quand
\di\la_+=-1_{\{\lceil x\rceil\mbox{ is odd}\}}\di x,
\ee
i.e., the measure $\di\la_-$ has a density with respect to the Lebesgue
measure which is $-1$ on the intervals $(1,2]$, $(3,4]$,\ldots and zero
elsewhere, and likewise, the density of $\di\la_+$ is $+1$ on the intervals
$(0,1]$, $(2,3]$,\ldots and zero elsewhere. Let $(\Xc_t)_{t\geq 0}$ denote a
model with such demand and supply functions (which satisfy (A1)--(A4)) and let
$\Xc'_t:=\Xc_t\circ\psi^{-1}$ denote the image of the measure $\Xc_t$ under
the map
\be
\psi(x):=\lceil x/2\rceil\qquad(x\in I).
\ee
Then $(\Xc'_t)_{t\geq 0}$ is a model in which limit orders can ony be placed
at discrete prices in $\{1,\ldots,n\}$. In particular, buy and sell limit
orders at prices that in the original model lie in an interval of the form
$(2(k-1),2k)$ are placed in such a way that they always match, with buy orders
on the right of sell orders. After applying the map $\psi$ all these orders are
mapped to the price $k$, i.e., they still match.

\subsection{History of the model}\label{S:hist}

The first reference for a model of the type we have just described is Stigler
\cite{Sti64}, who simulated a model with $\la_\pm(I_\mp)=0$ and $\rho=0$ where
$-\di\la_-$ and $\di\la_+$ are the uniform distributions on a set of 10
prices. Luckock \cite{Luc03} (who was apparently unaware of Stigler's work)
considered the general model with demand and supply functions satisfying
$\la_\pm(I_\mp)=0$ and with $\rho=0$. Assuming a special sort of stationarity,
Luckock was able to find explicit expressions for the equilibrium distribution
of the bid and ask prices of his model. In \cite{Pla11}, the model was once
again independently reinvented, this time with $-\di\la_-$ and $\di\la_+$ the
uniform distributions on a set of 100 prices. Building on this and Luckock's
work, models with $\la_\pm(I_\mp)>0$ were considered in \cite{Swa16}, who was
able to give a precise criterion for the positive recurrence of such
models. In the meantime, Yudovina \cite{Yud12a,Yud12b}, who was unaware of the
previous references, in her Ph.D.\ thesis considered the model for a general
class of demand and supply functions (though less general than those of
Luckock) and also introduced a construction involving infinite piles of limit
orders that is mathematically equivalent to setting
$\la_\pm(I_\mp)>0$. Together with Kelly \cite{KY16}, under certain technical
conditions, they were able to prove that the limit inferior and limit superior
as time tends to infinity of the bid and ask prices have certain deterministic
values, that they were able to calculate explicitly.

A characteristic feature of the Stigler-Luckock model is that buy
and sell orders arrive at a rate that is independent of the current price. By
contrast, a number of authors have considered models where limit orders are
placed at rates that are relative to the price of the last transaction
\cite{Mas00} or the opposite best quote \cite{CST10,SRR16}. A very general
but rather complicated model is formulated in \cite{Smi12}. See also
\cite{CTPA11} and chapter~4 of \cite{Sla13} for a (partial) overview of the
literature up to that point. Several authors also allow cancellation of orders.

In real markets, much of the trade seems to come from traders who speculate on
the price going up or down. In view of this, a model where orders are placed
relative to the current price may appear more realistic than the model we are
interested in. Nevertheless, for an asset to be interesting for traders, there
must always be some real demand and supply in the background, no matter how
much this may be obscured by other effects. An unrealistic aspect of our model
is that even traders who have a genuine interest in the asset and are not
merely speculating will usually not place limit orders very far from the
current price, but rather wait until the price reaches a level that is
acceptable to them. Thus, limit orders that are visibly written into the order
book in our model may in reality not be visible, although they are in a sense
still there in the form of traders silently waiting for the price to go up
or down.

The impossibility to cancel a limit order is surely an unrealistic aspect of
the Stigler-Luckcock model, that moreover greatly affects its long-time
behavior. Nevertheless, neglecting cancellation of orders may not be too bad
on intermediate time scales. Thus, the stationary behavior of the model may be
thought of as an idealization of the quasi-stationary behavior of real markets
on time scales when the number of orders is already large but cancellation is
not yet an important aspect of the market.\footnote{If in the dynamics of the
  Stigler-Luckock model, one replaces the infinite lifetime of limit orders by
  an exponential one, then the model becomes positive recurrent for any value
  of the parameters and the competitive window (see Section~\ref{S:window})
  becomes ill-defined. Nevertheless, as long as the cancellation rate is small
  compared to the arrival rate of orders, the quasi-stationary behavior of
  such a model is well approximated by a model without cancellation, and the
  competitive window can be understood in a limiting sense.}

In the days before electronic trading, market makers on the floor of the
exchange would match buy and sell orders. Even though nowadays, market makers
are not formally separated from other traders, they still exist in the form of
liquidity suppliers that are distinguished from other traders by having a
different motivation for trading. Rather than being interested in buying or
selling an asset or speculating on the future development of its price, market
makers place both buy and sell orders, at a high volume, with the aim of
profiting from the small difference between the bid and ask prices. The
strategy we have chosen for market makers is extremely simple. Depending on
the current state of the order book and the expected behavior of the other
traders, more intelligent choices may be possible. We will see, however, that
the presence of market makers in itself has a huge effect on the shape of the
order book. After this is taken into account, their present strategy may prove
not to be too unrealistic.

From a purely mathematical prespective, the Stigler-Luckock model is similar
to a number of other models that are motivated by other applications. We
mention in particular the Bak Sneppen model \cite{BS93} and its modification
by Meester and Sarkar \cite{MS12}, a model for canyon formation \cite{Swa15},
as well as the queueing models for email communication of Barab\'asi
\cite{Bar05} and Gabrielli and Caldarelli \cite{CG09}. All these models are
``rank-based'' in the sense that the dynamics are based on the relative order
of the particles and all models contain some rule of the form ``kill the
lowest (or highest) particle''. For the model of \cite{CG09}, the shape of the
stationary process near the critical point has been studied in \cite{FS15} and
these authors conjecture that their results also hold for the Stigler-Luckock
model.

\section{Behavior of the model without market makers}\label{S:orig}

\subsection{The competitive window}\label{S:window}

Consider a Stigler-Luckock model with $\la_\pm(I_\mp)=0$
and without market makers (i.e., $\rho=0$). Assumptions
(A1)--(A4) imply that there exists a unique price $x_{\rm W}\in I$ and
constant $V_{\rm W}>0$ such that \be \la_-(x_{\rm W})=\la_+(x_{\rm W})=:V_{\rm
  W}.  \ee Classical economic theory going back to Walras \cite{Wal74} says
that in a perfectly liquid market in equilibrium, a commodity with demand
and supply functions $\la_\pm$ is traded at the price $x_{\rm W}$ and the
volume of trade is given by $V_{\rm W}$. We call $x_{\rm W}$ the
\emph{Walrasian price} and $V_{\rm W}$ the \emph{Walrasian volume of trade}.

\begin{figure}[t]
\begin{center}
\begin{tikzpicture}[>=triangle 45,xscale=10,yscale=0.05]
\draw[very thin,fill=red,draw=white] plot file {P10000.dat};
\draw[very thin,fill=blue,draw=white] plot file {M10000.dat};
\draw[very thin] (0,0) -- (1,0);
\foreach \x in {0,0.2,0.4,0.6,0.8,1}
 \draw[very thick] (\x,0) -- (\x,-3) node[below] {\x};
\end{tikzpicture}
\caption{Simulation of the ``uniform'' Stigler-Luckock model with $\ov I=[0,1]$,
  $\la_-(x)=1-x$, and $\la_+(x)=x$. Shown is the state of the order book after
  the arrival of $10,000$ traders (starting from an empty order book).}
\label{fig:Luck}
\end{center}
\end{figure}
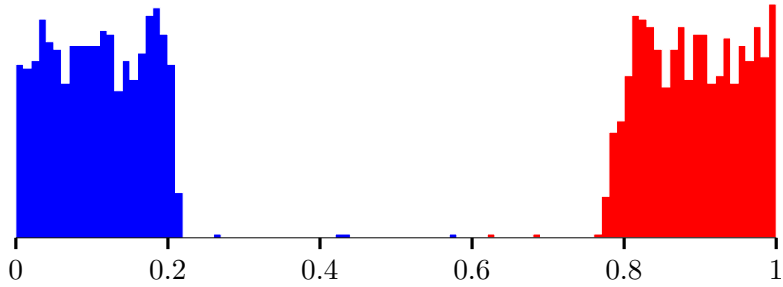

Perhaps not surprisingly, in the absence of market makers, Stigler-Luckock
models turn out to be highly non-liquid. Indeed, buyers willing to pay a
price above the Walrasian price $x_{\rm W}$ and sellers willing to sell for a
price below $x_{\rm W}$ may have to wait a considerable time before they get
their trade, since the bid and ask prices do not settle at $x_{\rm W}$ but
instead keep fluctuating in a \emph{competitive window} $(x_-,x_+)$ which
satisfies $\la_-(x_-)=\la_+(x_+)$. As a result, \emph{Luckock's volume of
  trade} $V_{\rm L}:=\la_-(x_-)=\la_+(x_+)$ is larger than the Walrasian
volume of trade $V_{\rm W}$ and in fact larger than it could be at any fixed
price level.

Figure~\ref{fig:Luck} shows the result of a numerical simulation of the
uniform model with $\ov I=[0,1]$, $\la_-(x)=1-x$, and $\la_+(x)=x$. Depicted
is the state of the order book, started from the empty initial state, after
the arrival of 10,000 traders.  This and more precise simulations suggest that
the boundaries of the competitive window are given by $x_-\approx0.218$ and
$x_+\approx0.782$. In the long run, buy limit orders at prices below $x_-$ and
sell limit orders at prices above $x_+$ stay in the order book forever, while
all other orders are eventually matched. As a result, Luckcock's volume of
trade $V_{\rm L}\approx0.782$ is considerably higher than the Walrasian volume
of trade $V_{\rm W}=0.5$. Luckock \cite{Luc03} described a method how to
calculate $x_-$, $x_+$, and $V_{\rm L}$. In particular, for the uniform
model, his method predicts that $V_{\rm L}=1/z$ with $z$ the unique solution
of the equation $e^{-z}-z+1=0$. To explain Luckock's formula for $V_{\rm L}$,
we need to look at restricted models.

\subsection{Restricted models}

Let $(\Xc_t)_{t\geq 0}$ be a Stigler-Luckock model defined by demand and
supply functions $\la_\pm:\ov I\to\R$ and rate of market makers $\rho\geq
0$. Let $(J_-,J_+)=J\sub I$ be an open subinterval of $I$ and let $\la'_\pm:\ov
J\to\R$ be the restrictions of the functions $\la_\pm$ to $J$. Let
$(\Xc'_t)_{t\geq 0}$ be the Stigler-Luckock model on $J$ defined by the by the
demand and supply functions $\la'_\pm$ and the rate of market makers
$\rho$. We call $(\Xc'_t)_{t\geq 0}$ the \emph{restricted model on} $J$. Its
dynamics are the same as for the original model $(\Xc_t)_{t\geq 0}$, except
that limit orders arriving outside $J$ cannot be written into the order
book. Instead, buy limit orders arriving at prices in $[J_+,I_+)$ are
converted into buy market orders while buy limit orders arriving at
prices in $(I_-,J_-]$ have no effect. Similar rules apply to sell limit
orders. Note that as long as the bid and ask prices $M^\pm_t$ stay inside $J$,
the evolution of both models inside $J$ is the same, i.e., restricting the
measure $\Xc_t$ to $J$ yields~$\Xc'_t$.

Consider, in particular, a Stigler-Luckock model with
$\la_\pm(I_\mp)=0$ and without market makers (i.e., $\rho=0$).
Let $\la^{-1}_-:[0,\la_-(I_-)]\to\ov I$ and $\la^{-1}_+:[0,\la_+(I_+)]\to\ov
I$ denote the left-continuous inverses of the functions $\la_-$
and $\la_+$, respectively, i.e.,
\be\label{jpm}
\la^{-1}_-(V):=\sup\{x\in\ov I:\la_-(x)\geq V\}
\quand
\la^{-1}_+(V):=\inf\{x\in\ov I:\la_+(x)\geq V\}.
\ee
Let $V_{\rm max}:=\la_-(I_-)\wedge\la_+(I_+)$ denote the maximal possible
volume of trade. To avoid trivialities, let us assume that
\begin{itemize}
\item[{\rm(A5)}] $V_{\rm W}<V_{\rm max}$.
\end{itemize}
By the continuity of the demand and supply functions, for
each $V\in(V_{\rm W},V_{\rm max}]$, setting $J(V):=(\la^{-1}_-(V),\la^{-1}_+(V))$
defines a subinterval $J(V)\sub I$ such that $\la_-(J_-(V))=V=\la_+(J_+(V))$.
For later use, we define a continuous, strictly increasing function
$\Phi:[V_{\rm L},V_{\rm max}]\to\R$ with $\Phi(0)=0$ by
\be\label{Phidef}
\Phi(V):=\int_{V_{\rm W}}^V\Big\{\frac{1}{\la_+\big(\la^{-1}_-(W)\big)}
+\frac{1}{\la_-\big(\la^{-1}_+(W)\big)}\Big\}\frac{1}{W^2}\di W.
\ee
By definition, a Stigler-Luckock model is \emph{positive recurrent} if started
from an empty order book, it returns to the empty state in finite expected
time. The following facts have been proved in \cite{Swa16}.

\bp[Luckock's volume of trade]
Assume\label{P:VLdef} (A1)--(A5), $\la_\pm(I_\mp)=0$ and
$\rho=0$. Then, for each $V\in(V_{\rm W},V_{\rm max})$, the restricted
Stigler-Luckock model on $J(V)$ is positive recurrent if and only if
$\Phi(V)<1/V_{\rm W}^2$.
\ep
\bpro
This follows from Proposition~2, Theorem~3, and formula (1.22) in
\cite{Swa16}.
\epro

Proposition~\ref{P:VLdef} suggests that Luckcock's volume of trade
should be given by
\be\label{VLdef}
V_{\rm L}=\sup\big\{V\in[V_{\rm W},V_{\rm max}):\Phi(V)<1/V_{\rm W}^2\big\},
\ee
and that the competitive window is given by $(x_-,x_+)=J(V_{\rm
  L})=\big(\la^{-1}_-(V_{\rm L}),\la^{-1}_+(V_{\rm L})\big)$. These formulas
agree well with numerical simulations and also agree with the (somewhat more
complicated) method for calculating $V_{\rm L}$ described in \cite{Luc03}. For
the uniform model, one can check that one obtains for $V_{\rm L}$ the constant
described at the end of the previous subsection. Under certain additional
technical assumptions on $\la_{\pm}$, which include the uniform model, it has
been proved in \cite[Thms~2.1 and 2.2]{KY16} that the limit inferior and limit
superior of the bid and ask prices are a.s.\ given by the boundaries of the
competitive window, as we have just calculated it.

We note that $V_{\rm L}>V_{\rm W}$ always but it is possible that $V_{\rm L}=V_{\rm
  max}$. In the latter case, the competitive window is the whole interval
$I$. For example, this happens for the model with $\ov I=[0,1]$,
$\la_-(x)=(1-x)^\al$, and $\la_+(x)=x^\al$ if $0<\al\leq1/2$. In the next
subsection, we will see that if $V_{\rm L}<V_{\rm max}$ and one assumes that
the restricted model on the competitive window has an invariant law, then the
equilibrium distributions of the bid and ask prices are given by the unique
solutions of a certain differential equation.

\subsection{Stationary models}

By definition, an invariant law for a Stigler-Luckock model is a
probability law on $\Si_{\rm ord}$ so that the process started in this initial
law is stationary. We let
\be
\Si^{\rm fin}_{\rm ord}:=\big\{\Xc\in\Si_{\rm ord}:\Xc^-\mbox{ and }\Xc^+
\mbox{ are finite measures}\big\}
\ee
denote the subspace of $\Si_{\rm ord}$ consisting of all states in which the
order book contains only finitely many orders. If a Stigler-Luckock model is
positive recurrent, then it has a unique invariant law that is moreover
concentrated on $\Si^{\rm fin}_{\rm ord}$ (see \cite[Thm~3]{Swa16}). In
particular, this applies to the restricted model on $J(V)$ for any $V<V_{\rm
  L}$. If $V_{\rm L}<V_{\rm max}$, then it is believed that the restricted
model on the competitive window $J(V_{\rm L})$ also has a unique invariant
law, but this invariant law is not concentrated on $\Si^{\rm fin}_{\rm
  ord}$. Instead, in equilibrium, the competitive window contains
a.s.\ infinitely many limit orders of each type. In \cite{FS15}, a precise
conjecture is made about the asymptotics of $\Xc^-$ near $J_-(V_{\rm L})$ and
$\Xc^+$ near $J_+(V_{\rm L})$ in equilibrium.

On a rigorous level, even just proving existence of an invariant law for the
restricted model on $J(V_{\rm L})$ is so far an open problem. However,
postulating the existence of such an invariant law, Luckock was able to
calculate the equilibrium distribution of the bid and as prices. We cite the
following result from \cite[Thm~1]{Swa16}. Essentially, this goes back to
\cite[formulas (20) and (21)]{Luc03}, although he only considers the case
$\la_\pm(I_\mp)=0$.

\bt[Luckock's differential equation]
Assume\label{T:Luckock} that a Stigler-Luckock model with demand and supply
functions satisfying (A1)--(A4) and $\rho=0$ has an invariant law. Let
$(\Xc_t)_{t\geq 0}$ denote the process started in this invariant law, and let
$M^\pm_t=M^\pm(\Xc_t)$ denote the bid and ask price at time $t\geq 0$. Define
functions $f_\pm:\ov I\to\R$ by
\be\label{fpm}
f_-(x):=\P\big[M^-_t\leq x\big]
\quand
f_+(x):=\P\big[M^+_t\geq x\big]\qquad(x\in\ov I),
\ee
which by stationarity do not depend on $t\geq 0$. Then $f_\pm$
are continuous and solve the equations
\be\ba{rr@{\,}c@{\,}l}\label{Luckock}
{\rm(i)}&\dis f_-\di\la_++\la_-\di f_+&=&\dis0,\\[5pt]
{\rm(ii)}&\dis f_+\di\la_-+\la_+\di f_-&=&\dis0,\\[5pt]
{\rm(iii)}&\multicolumn{3}{c}{\dis f_-(I_+)=1=f_+(I_-),}
\ec
where $f_-\di\la_+$ denotes the measure $\di\la_+$ weighted with the density
$f_-$, and the other terms have a similar interpretation.
\et

Consider a Stigler-Luckock model satisfying (A1)--(A5), $\la_\pm(I_\mp)=0$ and
$\rho=0$. Let $J$ be a subinterval such that $\ov J\sub I$.
Then it has been shown in \cite[Prop.~2]{Swa16} that Luckock's equation
(\ref{Luckock}) for the restricted model $(\Xc'_t)_{t\geq
  0}$ on $J$ has a unique solution $(f_-,f_+)$. By Theorem~\ref{T:Luckock},
if the restricted model on $J$ has an invariant law, then
\be
f_-(J_-)=\P\big[\Xc^{'\,-}_t=0\big]
\quand
f_+(J_+)=\P\big[\Xc^{'\,+}_t=0\big]
\ee
are the equilibrium probabilities that the restricted model $(\Xc'_t)_{t\geq
  0}$ contains no buy or sell limit orders, respectively. In particular, if
the restricted model on $J$ has an invariant law, then these quantities must
be $\geq 0$, and if the restricted model is positive recurrent they must be
$>0$. In \cite[Thm~3]{Swa16} it is shown that conversely, if $f_-(J_-)\wedge
f_+(J_+)>0$, then the restricted model on $J$ is positive recurrent. For
intervals of the form $J(V)=\big(\la^{-1}_-(V),\la^{-1}_+(V)\big)$ as in
(\ref{jpm}), it is shown in \cite[Prop~2 and formula (1.22)]{Swa16} that 
\begin{itemize}
\item If $\Phi(V)<1/V_{\rm W}^2$, then $f_-(\la^{-1}_-(V))>0$ and $f_+(\la^{-1}_+(V))>0$.
\item If $\Phi(V)=1/V_{\rm W}^2$, then $f_-(\la^{-1}_-(V))=0=f_+(\la^{-1}_+(V))$.
\end{itemize}
(Here $\Phi$ is the function defined in (\ref{Phidef}).) In particular, if
$V_{\rm L}<V_{\rm max}$, then Luckock's equation has a unique solution
$(f_-,f_+)$ on the competitive window $J(V_{\rm L})$, and this solution
satisfies $f_-(J_-(V_{\rm L}))=0=f_+(J_+(V_{\rm L}))$, which indicates that
the bid and ask prices never leave the competitive window.

\section{Behavior of the model with market makers}\label{S:market}

\subsection{Numerical simulation}

\begin{figure}[t]
\begin{center}
\begin{tikzpicture}[>=triangle 45,xscale=10,yscale=0.09]
\begin{scope}[xscale=0.45,yscale=6,xshift=0cm,yshift=-5cm]
\draw[very thin,fill=red,draw=white] plot file {Phistorho6a.dat};
\draw[very thin,fill=blue,draw=white] plot file {Mhistorho6a.dat};
\draw[very thin] (0,0) -- (1,0);
\foreach \x in {0,0.2,0.4,0.6,0.8,1}
 \draw[very thick] (\x,0) -- (\x,-0.15) node[below] {\x};
\draw (0.4,2) node {$\rho=0.6$};
\end{scope}

\begin{scope}[xscale=0.45,yscale=6,xshift=1.15cm,yshift=-5cm]
\draw[very thin,fill=red,draw=white] plot file {Phistorho6b.dat};
\draw[very thin,fill=blue,draw=white] plot file {Mhistorho6b.dat};
\draw[very thin] (0,0) -- (1,0);
\foreach \x in {0,0.2,0.4,0.6,0.8,1}
 \draw[very thick] (\x,0) -- (\x,-0.15) node[below] {\x};
\draw (0.4,2) node {$\rho=0.6$};
\end{scope}

\begin{scope}[xscale=0.45,yscale=6,xshift=2.3cm,yshift=-5cm]
\draw[very thin,fill=red,draw=white] plot file {Phistorho6c.dat};
\draw[very thin,fill=blue,draw=white] plot file {Mhistorho6c.dat};
\draw[very thin] (0,0) -- (1,0);
\foreach \x in {0,0.2,0.4,0.6,0.8,1}
 \draw[very thick] (\x,0) -- (\x,-0.15) node[below] {\x};
\draw (0.6,2) node {$\rho=0.6$};
\end{scope}

\begin{scope}[xscale=0.45,yscale=0.3,xshift=0cm]
\draw[very thin,fill=red,draw=white] plot file {Phistorho0.dat};
\draw[very thin,fill=blue,draw=white] plot file {Mhistorho0.dat};
\draw[very thin] (0,0) -- (1,0);
\foreach \x in {0,0.2,0.4,0.6,0.8,1}
 \draw[very thick] (\x,0) -- (\x,-4) node[below] {\x};
\draw (0.5,60) node {$\rho=0$};
\end{scope}

\begin{scope}[xscale=0.45,yscale=0.3,xshift=1.15cm]
\draw[very thin,fill=red,draw=white] plot file {Phistorho2.dat};
\draw[very thin,fill=blue,draw=white] plot file {Mhistorho2.dat};
\draw[very thin] (0,0) -- (1,0);
\foreach \x in {0,0.2,0.4,0.6,0.8,1}
 \draw[very thick] (\x,0) -- (\x,-4) node[below] {\x};
\draw (0.5,60) node {$\rho=0.2$};
\end{scope}

\begin{scope}[xscale=0.45,yscale=0.3,xshift=2.3cm]
\draw[very thin,fill=red,draw=white] plot file {Phistorho5.dat};
\draw[very thin,fill=blue,draw=white] plot file {Mhistorho5.dat};
\draw[very thin] (0,0) -- (1,0);
\foreach \x in {0,0.2,0.4,0.6,0.8,1}
 \draw[very thick] (\x,0) -- (\x,-4) node[below] {\x};
\draw (0.65,60) node {$\rho=0.5$};
\end{scope}
\end{tikzpicture}
\caption{Simulation of the uniform Stigler-Luckock model of
  Figure~\ref{fig:Luck} for different values of the rate $\rho$ of market
  makers. Shown is the state of the order book after the arrival of $10,000$
  traders. The histograms for $\rho=0.6$ have a different vertical scale.}
\label{fig:market}
\end{center}
\end{figure}
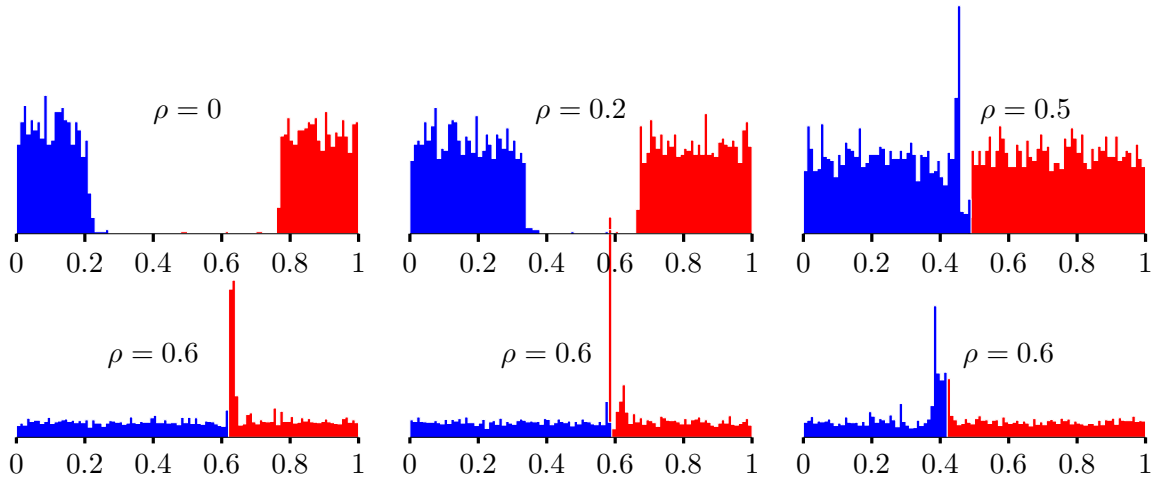

In Figure~\ref{fig:market}, we show the results of numerical simulations of
the ``uniform'' Stigler-Luckock model with $\ov I=[0,1]$, $\la_-(x)=1-x$, and
$\la_+(x)=x$, for different rates $\rho$ of market makers. We observe that as
$\rho$ is increased, the size of the competitive window decreases, until for
$\rho=\rho_{\rm c}=0.5$, it closes completely and the bid and ask prices
settle at the Walrasian price $x_{\rm W}$. If the rate $\rho$ of market makers
is increased even more beyond this point, we observe an interesting
phenomenon. In this regime, the bid and ask prices converge to a random limit
which is different each time we run the simulation, and which in
general also differs from the Walrasian price $x_{\rm W}$. The reason for this
is a huge surplus of limit buy and sell orders placed by market makers on the
current bid and ask prices, which is capable of ``freezing'' the price at a
random position.

In the coming subsections, we will demonstrate that the critical rate
$\rho_{\rm c}$ of market makers for which the competitive window closes
completely is for continuous models given by $\rho_{\rm c}=V_{\rm W}$, the
Walrasian volume of trade. We will argue that for $\rho<V_{\rm W}$, the
equilibrium distributions of the bid and ask prices are still given by the
unique solutions of a differential equation, similar to the one for the model
with $\la_\pm(I_\mp)=0$. For $\rho\geq V_{\rm W}$, we will prove that the bid
and ask price converge to a common limit and determine the subinterval of
possible prices where this limit can take values.

\subsection{Stationary models}\label{S:station}

In the present subsection, we show how for $0<\rho<V_{\rm W}$, one can
calculate the competitive window and the equilibrium distributions of the bid
and ask prices by methods similar to those for $\rho=0$. We first investigate
how Luckock's differential equation changes in the presence of market makers.

\bt[Luckock's differential equation]
Theorem~\ref{T:Luckock} \label{T:Luckock2} generalizes to $\rho\geq 0$
provided we modify Luckock's equation (\ref{Luckock}) to
\be\ba{rr@{\,}c@{\,}l}\label{Luckock2}
{\rm(i)}&\dis f_-\di\la_++(\la_--\rho)\di f_+&=&\dis0,\\[5pt]
{\rm(ii)}&\dis f_+\di\la_-+(\la_+-\rho)\di f_-&=&\dis0,\\[5pt]
{\rm(iii)}&\multicolumn{3}{c}{\dis f_-(I_+)=1=f_+(I_-).}
\ec
\et

\bpro
We first show that $f_\pm$ are continuous. By symmetry, it suffices to do this
for $f_-$. Right continuity is immediate from the continuity of the
probability measure $\P$. To prove continuity, it suffices to prove that
$\P[M^-_t=x]=0$ for all $x\in(I_-,I_+]$. This is clear for $x=I_+$. Imagine
that $\P[M^-_0=x]>0$ for some $x\in(I_-,I_+)$. Since $\Xc_0\in\Si_{\rm ord}$,
there are initially finitely many buy limit orders in $[x,I_+)$. By assumption
(A4), there is a positive probability that these buy limit orders are all
removed at some time before time one, while by assumption (A2), the
probability of a new buy limit order arriving at $x$ after such a time is
zero. This proves that $\P[M^-_1=x]<\P[M^-_0=x]$, contradicting stationarity.

To prove (\ref{Luckock2}), we observe that by stationarity, for each measurable
$A\sub I$ that is bounded away from $I_-$, sell limit orders are added in $A$
at the same rate as they are removed. This yields the equation
\be\label{fi}
\int_A\P[M^-_t<x]\,\di\la_+(\di x)+\rho\int_A\P[M^+_t\in\di x]
=\int_A\la_-(x)\,\P[M^+_t\in\di x].
\ee
Here, the first term on the left-hand side is the frequency at which sell
limit orders are added at a price $x\in A$ while the current bid price is
lower than $x$, the second term on the left-hand side is the frequency at
which market makers add sell limit orders at the current ask price, and the
right-hand side is the frequency at which sell limit orders at the current ask
price are removed because of the arrival of a buy limit order at a lower price
or the arrival of a buy market order. Using also continuity of $f_-$,
(\ref{fi}) proves (\ref{Luckock2})~(i). The proof of (ii) is similar while the
boundary conditions (iii) follow from the fact that $M^-_t<I_+$ and
$M^+_t>I_-$ a.s.
\epro

Assume (A1)--(A5), fix $\rho$ and define $\ti\la_\pm:=\la_\pm-\rho$. Then
$\di\ti\la_\pm=\di\la_\pm$ and hence (\ref{Luckock2}) is just Luckock's
original equation (\ref{Luckock}) with $\la_\pm$ replaced by $\ti\la_\pm$.
In particular, if $\rho<V_{\rm W}$, then
\be\label{tiVmax}
\ti V_{\rm max}:=\sup\big\{V\geq V_{\rm
  W}:\ti\la_-(\la^{-1}_+(V))\wedge\ti\la_+(\la^{-1}_-(V))>0\big\}
\ee
satisfies $V_{\rm W}<\ti V_{\rm max}$, and for each $V\in(V_{\rm W},\ti V_{\rm
  max})$, the functions $\ti\la_\pm$ are positive on the subinterval
$J(V)=(\la^{-1}_-(V),\la^{-1}_+(V))$. This suggests that for the model with
market makers, Luckock's volume of trade should be given by (\ref{VLdef}) but
with $V_{\rm max}$ replaced by $\ti V_{\rm max}$ and with the functions
$\la_\pm$ in the definition of $\Phi$ in (\ref{Phidef}) replaced by
$\ti\la_\pm$.

Defining $V_{\rm L}$ by this formula, if $V_{\rm L}<\ti V_{\rm max}$, then
\cite[Prop.~2]{Swa16} tells us that (\ref{Luckock2}) has a unique solution
$(f_-,f_+)$ on the competitive window $J(V_{\rm L})=(\la^{-1}_-(V_{\rm
  L}),\la^{-1}_+(V_{\rm L}))$, which should give the equilibrium distribution
of the bid and ask prices. Moreover, since $\ti V_{\rm max}$ (which depends on
$\rho$) decreases to $V_{\rm W}$ as $\rho\up V_{\rm W}$, we see that $V_{\rm
  L}\down V_{\rm W}$ and the size of the competitive window decreases to zero
as $\rho\up V_{\rm W}$.

\subsection{The regime with many market makers}\label{S:freeze}

In the previous subsection, we have argued that the competitive window has a
positive length for each $\rho<V_{\rm W}$ but its length decreases to zero as
$\rho\up V_{\rm W}$. In the present subsection, we look at the regime
$\rho\geq V_{\rm W}$. It will be necessary to strengthen assumptions (A1) and
(A3) on the demand and supply functions $\la_\pm$, to:
\begin{itemize}
\item[{\rm(A6)}] $\la_-$ is strictly decreasing on $I$ and $\la_+$ is strictly
  increasing on $I$.
\end{itemize}
We have argued in Subsection~\ref{S:model} that the assumptions (A1)--(A3) can
basically be made without loss of generality. Moreover, (A4) and (A5) only
exclude trivial cases. Assumption (A6) is restrictive, however. As explained
at the end of Subsection~\ref{S:model}, we can include models where prices
assume only discrete values in our analysis by constructing such models as
functions of other models which satisfy (A1)--(A3). However, as is clear from
(\ref{evenodd}), these models will not satisfy (A6), so our result
Theorem~\ref{T:freeze} below does not apply to discrete models.

For models with $\la_\pm(I_\mp)=0$, we generalize our previous definition of the
Walrasian volume of trade $V_{\rm W}$ by setting
\be
V_W:=\sup_{x\in\ov I}\big(\la_-(x)\wedge\la_+(x)\big).
\ee
Under the assumptions (A2) and (A6), the function $\la_-\wedge\la_+$ assumes
its maximum over $\ov I$ in a unique point $x_{\rm W}$, which we call the
Walrasian price. For models with $\la_\pm(I_\mp)=0$, these definitions agree
with our earlier definitions. The following theorem describes the behavior of
Stigler-Luckcock models with $\rho\geq V_{\rm W}$.

\bt[Fixation of the price]
Let\label{T:freeze} $(\Xc_t)_{t\geq 0}$ be a Stigler-Luckock model with demand
and supply functions $\la_\pm$ satisfying (A2), (A4), and (A6), and rate
of market makers $\rho$ satisfying $\rho\geq V_{\rm W}$, started in an
initial state in $\Si_{\rm ord}$. Let $M^\pm_t=M^\pm(\Xc_t)$ denote the bid
and ask price at time $t\geq 0$. Then there exists a random variable
$M_\infty$ such that
\be
\lim_{t\to\infty}M^-_t=\lim_{t\to\infty}M^+_t=M_\infty\quad{\rm a.s.}
\ee
Moreover, the support of the law of $M_\infty$ is given by $\{x\in\ov
I:\la_-(x)\vee\la_+(x)\leq\rho\}$. In particular, if $\rho=V_{\rm W}$, then
$M_\infty=x_{\rm W}$ a.s.
\et

We prepare for the proof of Theorem~\ref{T:freeze} with a number of lemmas,
some of which are of independent interest.

\bl[Lower bound on freezing probability]
Let\label{L:lofreez} $(\Xc_t)_{t\geq 0}$ be a Stigler-Luckock model on an
interval $I$ with demand and supply functions $\la_\pm$ satisfying (A1)--(A4)
and rate of market makers $\rho\geq 0$. Assume that initially $M^-_0=y$ where
$y\in I$ satisfies $\la_+(y)<\rho$. Then
\be
\P\big[M^-_t\geq y\ \forall t\geq 0\big]\geq 1-\frac{\la_+(y)}{\rho}.
\ee
\el

\bpro
Consider the number $\Xc^-_t(\{y\})$ of buy limit orders that are placed
exactly at the price $y$. At times when $M^-_t=y$, this quantity goes up by
one with rate $\rho$ and down by one with rate $\la_+(y)$, while at times when
$M^-_t>y$, this quantity does not change at all. Thus, up to the first time
that $\Xc^-_t(\{y\})=0$, this process is a random time change of the random
walk on $\Z$ that jumps up one step with rate $\rho$ and down one step with rate
$\la_+(y)$. If $\la_+(y)<\rho$, then by the well-known gambler's ruin, this
random walk, started in 1, stays positive with probability $1-\la_+(y)/\rho$.
\epro

\bl[Bound on the competitive window]
Let\label{L:nospread} $(\Xc_t)_{t\geq 0}$ be a Stigler-Luckock model on an
interval $I$ with demand and supply functions $\la_\pm$ satisfying (A1)--(A4)
and rate of market makers $\rho\geq 0$. Assume that $x,y\in I$ satisfy
$\la_-(x)>\la_-(y)$ and $\la_+(y)<\rho$. Then
\be\label{nospread}
\P\big[\liminf_{t\to\infty}M^-_t<x
\mbox{ and }
\limsup_{t\to\infty}M^+_t>y\big]=0.
\ee
By symmetry, the same conclusion can be drawn if $\la_+(x)<\la_+(y)$ and
$\la_-(x)<\rho$.
\el

\bpro
If we start the process in an intial state such that $M^+_0\geq y$, then there
is a probability
\be
p:=\frac{\la_-(x)-\la_-(y)}{\la_-(I_-)+\la_+(I_+)+\rho}>0
\ee
that the first trader arriving at the market places a buy limit order
somewhere in the interval $(x,y)$. By Lemma~\ref{L:lofreez}, there is then a 
probability of at least $q:=1-\la_+(y)/\rho>0$ that after this event, the best
buy price $M^-_t$ never drops to values $\leq x$ anymore. Thus, letting $\sig$
denote the first time that a trader arrives at the market, we have that
\be\label{lowpq}
\P\big[M^-_t>x\ \forall t\geq\sig\,|\,M^+_0\geq y\big]
\geq pq>0.
\ee
We claim that this implies (\ref{nospread}). To see this, set $\tau_0:=0$ and
define inductively
\be\ba{r@{\,}c@{\,}ll}
\dis\sig_k&:=&\dis\inf\{t\geq\tau_k:M^+_t\geq y\}\qquad&\dis(k\geq 0),\\[5pt]
\dis\sig'_k&:=&\dis\inf\{t>\sig_k:\mbox{ a trader arrives}\}
\qquad&\dis(k\geq 0),\\[5pt]
\dis\tau_k&:=&\dis\inf\{t\geq\sig'_{k-1}:M^-_t\leq x\}\qquad&\dis(k\geq 1),
\ec
where the infimum over the empty set is $:=\infty$. By the strong Markov
property, $P[\tau_k<\infty]\leq(1-pq)^k$ and hence $\P[\tau_k<\infty\ \forall
  k\geq 0]=0$, which implies (\ref{nospread}).
\epro

\bl[Freezing]
Let\label{L:freeze} $(\Xc_t)_{t\geq 0}$ be a Stigler-Luckock model with demand
and supply functions $\la_\pm$ satisfying (A2), (A4), and (A6), and rate
of market makers $\rho$ satisfying $\rho\geq V_{\rm W}$. Then there exists a
random variable $M_\infty$ such that
\be\label{freez}
\lim_{t\to\infty}M^-_t=\lim_{t\to\infty}M^+_t=M_\infty\quad{\rm a.s.}
\ee
\el

\bpro
If (\ref{freez}) does not hold, then there must exist $x,y\in I$ with $x<y$
such that
\be\label{poswind}
\P\big[\liminf_{t\to\infty}M^-_t<x
\mbox{ and }
\limsup_{t\to\infty}M^+_t>y\big]>0.
\ee
By (A6), making the interval $(x,y)$ smaller if necessary we can assume
without loss of generality that we are in one of the following two cases:
I.\ $\la_+(y)<\rho$, and II $\la_-(x)<\rho$. Using again (A6), we see that
(\ref{poswind}) contradicts Lemma~\ref{L:nospread}.
\epro

\bl[Bound on possible limit values]
Under\label{L:Mbd} the assumptions of Lemma~\ref{L:freeze}, the random
variable $M_\infty$ from (\ref{freez}) satisfies
\be
\la_-(M_\infty)\vee\la_+(M_\infty)\leq\rho\quad{\rm a.s.}
\ee
\el

\bpro
By symmetry, it suffices to prove that $\la_+(M_\infty)\leq\rho$ a.s. Assume
the converse. Then there exists some $z\in I$ with $\la_+(z)>\rho$ such that
$\P\big[M_\infty\in(z,I_+]\big]>0$. By the continuity of $\la_-$, for each
$\eps>0$, we can cover the compact interval $[z,I_+]$ with finitely many
intervals of the form $(x,y)$ (if $y<I_+$) or $(x,y]$ (if $y=I_+$) such that
$\la_-(x)-\la_-(y)\leq\eps$.  In view of this, we can find $x<y$ and $u>0$
such that $\la_+(x)>\rho+\big(\la_-(x)-\la_-(y)\big)$ and $\P[x\leq
 M^-_t\leq M^+_t\leq y\ \forall t\geq u]>0$.

During the time interval $[u,\infty)$, the number of buy limit orders in
$[x,y)$ can only increase when a market maker arrives or a buyer places a buy
limit order in $[x,y)$. On the other hand, the number of buy limit orders in
$[x,y)$ decreases each time a trader places a sell market order or a sell
limit order at some price in $(I_-,x]$, which happens at times according to a
Poisson process with rate $\la_+(x)$. Since
$\la_+(x)>\rho+\big(\la_-(x)-\la_-(y)\big)$, by the strong law of 
large numbers applied to the Poisson processes governing the arrival of
different sorts of traders, we see that a.s.\ on the event that $x\leq
 M^-_t\leq M^+_t\leq y\ \forall t\geq u$, there must come a time when there
 are no buy limit orders left in $[x,y)$, which is a contradiction.
\epro

\bpro[of Theorem~\ref{T:freeze}]
Lemmas~\ref{L:freeze} and \ref{L:Mbd} show that $M^\pm_t$ converge a.s.\ to a
common limit $M_\infty$ which takes values in the compact interval
$C:=\{x\in\ov I:\la_-(x)\vee\la_+(x)\leq\rho\}$. If $\rho=V_{\rm W}$, then by
(A6), $C$ consists of the single point $C=\{x_{\rm W}\}$. On the other hand,
if $\rho>V_{\rm W}$, then by (A6), $C=[C_-,C_+]$ is an interval of positive
length.  To complete the proof, we must show that in the latter case, for each
$C_-<x<y<C_+$, the event $M_\infty\in(x,y)$ has positive probability.  It is
not hard to see that for each $\Xc_0\in\Si_{\rm ord}$ and $t>0$, there is a
positive probability that $x<M^-_t<M^+_t<y$. Thus, it suffices to prove that
if $x<M^-_0<M^+_0<y$, then $\P[M_\infty\in(x,y)]>0$. This is similar to
Lemma~\ref{L:lofreez}, but we use a slightly different argument.

Note that by (A6), $\la_-(x)<\rho$ and $\la_+(y)<\rho$. As long as $x\leq
M^-_t\leq M^+_t\leq y$, the number $\Xc^-_t\big([x,y]\big)$ of buy limit
orders in $[x,y]$ goes up by one with rate at least $\rho$ and decreases by
one with rate at most $\la_+(y)$. A similar statement holds for the number of
sell limit orders in $(x,y)$. Let $(N^-_t,N^+_t)_{t\geq 0}$ be a Markov
process in $\Z^2$ that jumps with rates
\be\ba{l}
(n_-,n_+)\mapsto(n_-+1,n_+)\quad\mbox{at rate}\quad\rho,
\qquad\quad
(n_-,n_+)\mapsto(n_--1,n_+)\quad\mbox{at rate}\quad\la_+(y),\\[5pt]
(n_-,n_+)\mapsto(n_-,n_++1)\quad\mbox{at rate}\quad\rho,
\qquad\quad
(n_-,n_+)\mapsto(n_-,n_+-1)\quad\mbox{at rate}\quad\la_-(x).
\ec
Then $(N^-_t)_{t\geq 0}$ and $(N^+_t)_{t\geq 0}$ are independent random walks 
with positive drift, and hence by the strong law of large numbers, if
$N^-_0>0$ and $N^+_0>0$, then
\be
\P[N^-_t>0\mbox{ and }N^+_t>0\ \forall t\geq 0]>0.
\ee
The claim now follows from a simple coupling argument, comparing 
$\Xc^\pm_t\big([x,y]\big)$ with $N^\pm_t$.
\epro

\subsection{Conclusion}

The Stigler-Luckock model is one of the most basic and natural models for
traders interacting through a limit order book, so natural, in fact, that it
has been at least four times independently (re-)invented
\cite{Sti64,Luc03,Pla11,Yud12b}. Although it is based on natural assumptions,
its behavior is unrealistic since the bid and ask prices do not settle at the
Walrasian equilibrium price but rather keep fluctuating inside an interval of
positive length called the competitive window. This provides an opportunity
for market makers or liquidity suppliers who make money from buying at a low
price and selling at a higher price.

In this paper, we have added such market makers to the model who trade using a
very simple strategy, namely, by placing one buy and sell limit order at the
current bid and ask prices. We have seen that the addition of market makers
makes the model more realistic in the sense that the size of the competitive
window decreases. In particular, for continuous models, if the rate at which
market makers place orders equals the Walrasian volume of trade, then the size
of the competitive window decreases to zero and the bid and ask prices
converge to the Walrasian equilibrium price. If the rate of market makers is
even higher, then the bid and ask prices also converge to a common limit, but
now the limit price is random and in general differs from the Walrasian
equilibrium price. Moreover, in this regime, some of the limit orders placed
by market makers are never matched by market orders but stay in the order book
forever (on the time scale we are interested in).

In reality, market makers make profit only if their limit orders are matched,
and this profit is proportional to the size of the competitive
window. Therefore, in real markets, there is no motivation for market makers
to trade once the size of the competitive window has shrunk to zero. In view
of this, in reality, we can expect a self-regulating mechanism that makes sure
that in the long run, the rate at which market makers place orders is
approximately equal to the Walrasian volume of trade. The effect of this is
that in the limit, all trade involves market makers, i.e., the buyers and
sellers of the original Stigler-Luckock model do not directly interact with
each other but make all their trade with the market makers.

We conclude from this that adding market makers to the Stiger-Luckock model
has produced a more realistic model, especially if the rate of market makers
is chosen equal to the Walrasian volume of trade. Future, better models should
include a self-regulating mechanism that links the rate at which market makers
place orders to the present state of the order book by weighing their expected
profit against the costs and risks. Realistic models should also consider
prices that can take only discrete values since in reality the size of the
competitive window and hence the potential for profit for market makers are
bounded from below by the tick size.


\begin{thebibliography}{CTPA11}



\bibitem[Bar05]{Bar05}
A.-L.~Barab\'asi.
The origin of bursts and heavy tails in human dynamics.
\emph{Nature}~435 (2005), 207--211.

\bibitem[BS93]{BS93}
P.~Bak and K.~Sneppen.
Punctuated equilibrium and criticality in a simple model of evolution.
\emph{Phys.\ Rev.\ Lett.}~74 (1993), 4083--4086.



\bibitem[CG09]{CG09}
A.~Gabrielli and G.~Caldarelli.
Invasion percolation and the time scaling behavior of a queuing model of human
dynamics.
\emph{J.\ Stat.\ Mech.} (2009), P02046 (10 pages).

\bibitem[CST10]{CST10}
R.~Cont, S.~Stoikov, and R.~Talreja.
A stochastic model for order book dynamics.
\emph{Oper.\ Res.}~58(3) (2010), 549--563.

\bibitem[CTPA11]{CTPA11}
A.~Chakraborti, I.M.~Toke, M.~Patriarca, and F.~Abergel.
Econophysics review II: Agent-based models.
\emph{Quant. Finance}~11(7) (2011), 1013--1041.

\bibitem[FS16]{FS15}
M.~Formentin and J.M.~Swart.
The limiting shape of a full mailbox.
\emph{ALEA}~13(2) (2016), 1151--1164.




\bibitem[KY18]{KY16}
F.~Kelly and E.~Yudovina.
A Markov model of a limit order book: thresholds, recurrence,
and trading strategies.
\emph{Math.\ Oper.\ Res.}~43(1) (2018), 181--203.




\bibitem[Luc03]{Luc03}
H.~Luckock.
A steady-state model of the continuous double auction.
\emph{Quant. Finance}~3(5) (2003), 385--404.


\bibitem[Mas00]{Mas00}
S.~Maslov.
Simple model of a limit order-driven market
\emph{Physica A}~278 (2000), 571--578.

\bibitem[MS12]{MS12}
R.~Meester and A.~Sarkar.
Rigorous self-organised criticality in the modified Bak-Sneppen model.
\emph{J.\ Stat.\ Phys.}~149 (2012), 964--968.




\bibitem[Pla11]{Pla11}
Jana Pla\v{c}kov\'a.
\emph{Shluky volatility a dynamika popt\'avky a nab\'idky.} (In Czech)
Master Thesis, MFF, Charles University Prague, 2011.

\bibitem[Sla13]{Sla13}
F.~Slanina. 
\emph{Essentials of Econophysics Modelling.}
Oxford University Press, 2013.

\bibitem[Smi12]{Smi12}
M.~\v{S}m\'id.
Probabilistic properties of the continuous double auction.
\emph{Ky\-ber\-ne\-ti\-ka} 48(1) (2012), 50--82.

\bibitem[SRR17]{SRR16}
E.~Scalas, F.~Rapallo, and T.~Radivojevi\'c.
Low-traffic limit and first-passage times for a simple model of the continuous
double auction.
\emph{Phys.\ A}~485(1) (2017), 61--72.

\bibitem[Sti64]{Sti64}
G.J.~Stigler.
Public regulation of the securities markets.
\emph{The Journal of Business}~37(2) (1964), 117--142.


\bibitem[Swa17]{Swa15}
J.M.~Swart.
A simple rank-based Markov chain with self-organized criticality.
\emph{Markov Process.\ Related Fields}~23(1) (2017), 87--102.

\bibitem[Swa18]{Swa16}
J.M.~Swart.
Rigorous results for the Stigler-Luckock model for the evolution of an order
book.
\emph{Ann.\ Appl.\ Probab.}~28(3) (2018), 1491--1535.


\bibitem[Wal74]{Wal74}
L.~Walras.
\emph{\'El\'ements d' \'economie politique pure, ou Th\'eorie de la richesse
  sociale.}
First published 1874; published again in Paris by R.~Pichon and
R.~Durand-Auzias and in Lausanne by F.~Rouge, 1926.



\bibitem[Yud12a]{Yud12a}
E.~Yudovina.
A simple model of a limit order book.
Preprint (2012), ArXiv: 1205.7017v2

\bibitem[Yud12b]{Yud12b}
E.~Yudovina.
\emph{Collaborating Queues: Large Service Network and a Limit Order Book.}
Ph.D.\ thesis, University of Cambridge, 2012.

\end{thebibliography}
\end{document}